\documentclass[11pt,twoside]{article}

\usepackage{asp2006}
\usepackage{epsf}
\usepackage{lscape}

\begin{document}

\title{Kinematics of AGN and Quasar Jets}   

\author{K.~I.~Kellermann$^1$, M.~L.~Lister$^2$, D.~C.~Homan$^3$, Y.~Y.~Kovalev$^{4,5}$, M.~Kadler$^{6,4}$, \& M.~C.~Cohen$^7$} \affil{ $^1$National Radio Astronomy Observatory, 520 Edgemont Road, Charlottesville, VA 22903, U.S.A \\ $^2$Department of Physics, Purdue University, 525 Northwestern Avenue, West Lafayette, IN 47907, U.S.A.
\\
$^3$Department of Physics and Astronomy, Denison University, Granville, OH 43023, U.S.A.
\\
$^4$Max-Planck-Institut f\"ur Radioastronomie, Auf dem H\"ugel\,69,
53121 Bonn, Germany
\\
$^5$Astro Space Center of Lebedev Physical Institute, Profsoyuznaya 84/32, 117997 Moscow, Russia \\ $^6$Astrophysics Science Division, NASA's Goddard Space Flight Center, Greenbelt Road, Greenbelt, MD 20771, USA \\ $^7$Department of Astronomy, Mail Stop 105-24, California Institute of Technology, Pasadena, CA 91125, U.S.A.
}

\begin{abstract} 
The major multi-epoch VLBA programs are described and discussed in terms of relativistic beaming models.  Broadly speaking the observed kinematics are consistent with models having a parent population which is only mildly relativistic but with Lorentz factors extending up to about 30.  While the collimation and acceleration appears to mainly occur close to the central engine, there is evidence of accelerations up to 1 kpc downstream.  Generally the motion appears to be linear, but in some sources the motion follows a curved trajectory.  In other sources, successive features appear to be ejected in different directions possibly the result of a precessing nozzle.  

The launch of GLAST in 2008 will offer new opportunities to study the relation between radio and gamma--ray activity, and possibly to locate the source of the gamma--ray emission. VSOP-2 will give enhanced resolution and will facilitate the study of the two-dimensional structure of relativistic jets, while RadioAstron will provide unprecedented resolution to study the fine scale structure of the jet base.

\end{abstract}


\section{Introduction}   
As discussed earlier in this conference by Dave Meier, the theoretical framework for the formation of relativistic jets due to the accretion of matter onto a supermasive black hole is well developed.  Observations of jet kinematics obtained by multi-epoch high resolution imaging provides tests of theoretical models and can constrain the parameters of the jet formation.  However, the interpretation is not straightforward.  The observations do not give instantaneous multi-epoch snapshots of the source structure.  Rather, due to relativistic effects, different parts of the source are observed at different times, giving the illusion of superluminal motion, the distortion of angles and misrepresentation of luminosity.   In particular, the observed speeds and luminosities of jets are highly biased due to Doppler boosting within a narrow cone, and they are not representative of the intrinsic distributions.  Finally, we note that apparent speeds derived from multi-epoch VLBI observations refer to the observed motion of recognizable features, typically referred to as "components," which do not necessarily reflect the underlying bulk relativistic flow, but which instead may simply reflect the propagation of forward and reverse shocks within the jet flow.  The observations are further complicated since these features do not remain stable with time, but may break up into sub components, while other features may merge into a single complex feature e.g, \cite{K08}.

\section{The major multi-epoch programs}
Prior to the completion of the VLBA, observations of jet kinematics were based on infrequent multi-epoch VLBI observations using ad-hoc VLBI arrays generally with too few antennas to obtain the high-sensitivity and high dynamic range imaging needed to unambiguously trace jet motions.  A number of VLBA programs have now been underway for up to a decade or more at wavelengths between 7 mm and 13 cm and are providing new insight into jet kinematics.  This paper reports on global source kinematics primarily obtained from 2 cm VLBA observations, but we also discuss the multi-wavelength kinematics of a few particularly interesting sources.

\cite{B07} have combined their 6 cm multi-epoch observations of compact radio sources with pre-VLBA observations to study the kinematics of 266 radio sources.  Typically 2 to 4 epochs were observed for each source.

\citet{PFMG07} have used the Radio Reference Frame Image Database (RRFID) to determine source kinematics for 87 sources at 4 cm over the period 1994 to 1998.  This is part of a database originally obtained for precise astrometry and to study earth motions and includes multi-epoch observations at both 4 and 13 cm for up to 500 sources.
 
\citet{K04} and \citet{C07} have reported 2 cm VLBA studies for 110 AGN jets covering the period 1994 to 2001. Ros et al. and Lister et al. (in preparation) extend the observations to 2003 and 2007 respectively.  Since 2002, the MOJAVE program (Monitoring of Jets in Active Galactic Nuclei) has further extended this program to include the kinematics of a complete sample of the  the brightest 133 sources in the northern sky \citep{LH05} as well as 59 other sources of particular interest. Since 1994, these programs have observed as many as 30 epochs for some sources, permitting for the first time clear evidence for non radial motions, bends, accelerations and decelerations.  The MOJAVE program also includes linear \citep{LH05} and circular \citep{HL06} polarization measurements which can be used to investigate three-dimensional magnetic structure.  The MOJAVE program is a large VLBA program and images of new and past 2 cm VLBA observations are made available on the MOJAVE web site\footnote{http://www.physics.purdue.edu/MOJAVE} within a few weeks following the correlation of the data.
 
In a series of papers, the Boston University group has reported on 0.7 and 1.3 cm VLBA observations \citep{J05, J07}.\footnote{http://www.bu.edu/blazars/VLBAproject.html} These observations were made more frequently, typically monthly or bi-monthly, and consequently include a much smaller number of sources than observed by the longer wavelength programs. Monthly observations of 29 sources at 7 mm are continuing with special attention to radio counterparts of EGRET gamma-ray sources.

Recently TANAMI, (Tracking Active Galactic Nuclei with Australian-South-African Milliarcsecond Interferometry) a new VLBI  program to monitor extragalactic jets in a sample of sources below -30 degrees declination began to observe in November 2007 at 8.4\,GHz and 22\,GHz with the Australian South-African Long Baseline Array \citep{K08}. 
It is planned to expand this sample to include GLAST detected sources.

\section{Jet Properties} 
 
Due to relativistic beaming the observed brightness temperature is enhanced by the Doppler factor ($T_\mathrm{obs} = \delta \times T_\mathrm{int}$).  The maximum brightness temperature or lower limit that can be determined by VLBI depends only on flux density, measurement accuracy and baseline length, and is independent of the wavelength used.   See Kovalev (these proceedings).  For a source of $\sim$ 1 Jy and earth baselines, this limit is $\sim 10^{13}$~K.  Using the VLBA at 2 cm, \citet{Kov05} have observed brightness temperatures up to $\sim 5 \times 10^{13}$~K, while \citet{Hor04} have reported values of $T_b$ up to $\sim 10^{14}$~K using HALCA at 6 cm.  This suggests Doppler factors up to a few hundred if $T_\mathrm{int}$ is limited by Inverse Compton cooling to values  $\sim 10^{11-12}$~K.  However, the highest observed Lorentz factors are only 35 to 40 \citep{C07}, much less than required to explain the highest observed brightness temperatures, unless these high observed values of $T_\mathrm{obs}$ are only transient. 

\cite{H06} have compared the distribution of observed brightness temperatures with relativistic beaming models and find that in their medium to low brightness state, most sources have $T_\mathrm{int} \sim 3 \times 10^{10}$~K or close to an equipartition value.  However, when they are in a flaring state, $T_\mathrm{int} \geq 2 \times 10^{11}$~K,  which is close to the IC limit.  This correponds to an excess of particle over magnetic density of as much as $10^5$ or more.
  
\citet{K04} and \citet{J05} have used the characteristic variability time-scales to estimate intrinsic source size and brightness temperature, and find that the distribution of observed speeds and brightness temperature are consistent with $T_\mathrm{int}$ being close to the equilibrium value.  

\cite{C07} have shown that the
distribution of observed speed and luminosity is consistent with
relativistic beaming models having a maximum
Lorentz factor $\sim 32$, and intrinsic luminosity $\sim
10^{26}$~\,W\,Hz$^{-1}$.  Sources with low apparent radio luminosity and slow speeds are unlikely to be powerful highly relativistic jets oriented near the plane of the sky; nor can they be due to low luminosity high speed jets pointed close to the line of sight as the probability of these configurations is too low.

With the extensive data base now available, many jets show clear curvature.  A major question is to find out if the jet flow follows a preset curved trajectory or is ballistic.   Ballistic motion can produce a curved jet if it originates from a rotating nozzle, as might be induced if the AGN is part of a binary system. However, there is no clear model which explains all observations.  In some sources like $0738+313$, successive components appear to follow the same curved trajectory as previously ejected features \citep{L06}. However, in other sources, such as 1308+328, \citet{L06} has shown that the observed pattern is consistent with that expected from a precessing nozzle having a $6.3$ year period.


Although gamma-ray emission is associated with flat spectrum radio blazars, 
the limited sensitivity and poor time resolution of the existing EGRET observations has precluded a clear understanding of the relation between the radio and gamma--ray emission.  \cite{Kov05} found that gamma--ray blazars have  more compact structure than non gamma--ray blazars, while \citet{J01} and \citet{K04} reported somewhat faster observed speeds for gamma--ray detected sources.  \cite{J01} suggest that gamma--ray flares occur near the time of ejection of new superluminal components.

\section{Some sources of special interest}
M87 is one of the closest known relativistic jets and is of special interest since 1 mas corresponds to a linear scale of only 0.08 pc and an angular velocity of 1 mas/yr corresponds to a linear velocity of 0.25 c. As early as 1964, \citet{S64}
argued that the jet in M87 was intrinsically two-sided but appeared one-sided due to differential Doppler boosting of the approaching and receding relativistic jets.  This was the first suggestion that bulk relativistic motion might play an important role in understanding extragalactic radio sources.  However, curiously \citet{Kov07} find no evidence at 2 cm over a period of 12 years for any motions in any of seven apparent features greater than a few percent $c$ within about 1.5 pc of the jet base.  However, \cite{BSM99} report superluminal motion in the optical feature known as HST-1 located 0.86 arcsec (60 pc) downstream, while
\citet{CHS07} report $v/c \sim 4.5$ in the 20 cm radio feature associated with HST-1.  Also, \cite{A06} have reported strong and variable TeV emission from M87, and \cite{CHS07} argue that the TeV emission may originate in HST-1 and not at the base of the jet. Closer in to the base, 7 mm observations by \citet{LWJ07} 
with submilliarcsecond resolution suggests speeds which are a significant fraction of $c$. 

The high linear resolution 2 cm image of \citet{Kov07}
shows the beginning of the counter-jet. Kovelev et al. suggested that the apparent limb brightening of the main jet is due to a two component spine-sheath configuration as expected from theoretical models \citep{A01, H07}.  
In the context of these models, the 2 cm observations see only the slowly moving outer sheath, while the faster moving inner spine is responsible for the observed high energy emission.  However, \cite{TG08} speculate that the TeV emission might be produced in the slowly moving outer layer surrounding a fast misaligned spine.
 
3C 279 is one of the strongest radio blazers and its jet kinematics have been studied for more than 30 years.  \citet{H03} have reported an abrupt change in the trajectory of the brightest jet feature with an apparent change in direction of $\sim 26$ deg and corresponding velocity change from 8$c$ to 13$c$.  \citet{H03} concluded that the intrinsic bend was only about 1 degree, but occured more than 1 kpc away from the base of the jet, and that the focusing and collimation of relativistic jets can occur well downstream.

3C 111 is one of only a few known classical double lobed radio galaxies with observed superluminal motion in the core.  
\cite{K08}
have studied the evolution of the one sided radio jet in 3C 111, over a period of about 10 years during a major flare and jet-plasma ejection, and found a complex pattern of separating components possibly indicative of forward- and backward directed shocks as well as trailing components formed in the wake of the primary perturbation. Such effects may be interpreted as a result of coupling to Kelvin-Helmholtz instability pinching modes from the interaction of the jet with the external medium. They complicate the interpretation of multi-epoch VLBI jet observations and provide a challenge for the descriptive analysis of jet structure as well as for numerical jet models."

\section{Summary and New Opportunities}
Although the parent population of AGN jets is mostly only mildly relativistic, in flux density-limited samples, Doppler boosting causes the observed distribution to be highly biased toward the fastest jets with observed Lorentz factors up to 35 or 40 which are aligned close to the line-of-sight.  However, subluminal and stationary features are also observed.  Although each jet appears to have a characteristic velocity which may reflect the bulk flow, quantitative analysis may be complicated by differences between the bulk plasma flow and observed pattern motions, by the apparent splitting and merging of individual features, by finite opening angles, and by transverse velocity gradients.  On average quasars have larger Lorentz factors ($ \gamma \sim 8$) than low luminosity AGN, and EGRET detected blazars may have somewhat larger Lorentz factors.  The observed distribution of Lorentz factor and luminosity is consistent with Doppler boosting models, but there appear to be intrinsic differences as well as projection effects. Both ballistic motion characteristic of a rotating nozzle and curved trajectories with accelerations and decelerations are observed, and there is some evidence that in--situ acceleration along the jet may be important. The ejection of new features may be associated with radio or gamma--ray outbursts, but the evidence for this is inconclusive.  Brightness temperatures up to $10^{14}$~K are observed directly and are also inferred from the time scale of flux density variability.

Over the course of the next few years, the bandwidth of the VLBA will be increased giving an improvement in sensitivity of up to a factor of 5 for sustained observations.  In Europe, the EVN, especially when combined with eMERLIN, will give improved surface brightness sensitivity, and eVLBI will offer new target-of-opportunity possibilities, with real-time bandwidths of 1 Gbps.  Continued radio observations, combined with the unprecedented GLAST all sky monitoring, may give insight into the location and mechanism of gamma-ray outbursts.  In 2012, VSOP-2 will give an order of magnitude improvement in both sensitivity and resolution over HALCA along with polarization capability.  This will allow the transverse structures and magnetic field orientation in the M87 and other jets, to be studied in much better detail than previously possible.  In particular, VSOP-2 will better define the structure and evolution of jet-deflection/interaction surfaces and the role Kelvin-Helmholtz instabilities in sources such as 3C 111 and 3C 279.  However, frequent time sampling and good u,v coverage will be needed to fully exploit these capabilities.


\acknowledgements The National Radio Astronomy Observatory
is operated by Associated Universities, Inc., under a cooperative
agreement with the National Science Foundation.  YYK is currently a
Research Fellow of the Alexander von Humboldt Foundation.  
DCH was partially 
supported by an award from the Research Corporation.  MK was supported 
in part through a stipend from the International Max Planck Research School for Radio Astronomy at the University of Bonn and, in part, by a NASA Postdoctoral Program Fellowship
appointment at the Goddard Space Flight Center.  The MOJAVE
project is supported under National Science Foundation grant AST-0406923.

\end{document}